\documentclass[12pt,preprint]{aastex}

\shorttitle{Diagnostics of  3C 273 jet X-ray emission models}
\shortauthors{Georganopoulos, Perlman, Kazanas, \&  KcEnery}

\begin{document}

\title{Quasar X-ray jets: $\gamma$-ray diagnostics of the
synchrotron and inverse Compton hypotheses, the case of 3C 273}

\author{Markos Georganopoulos\altaffilmark{1,2}
Eric S. Perlman\altaffilmark{1} 
Demosthenes Kazanas\altaffilmark{2}
Julie McEnery\altaffilmark{2}}

\altaffiltext{1}{Department of Physics, Joint Center for Astrophysics, University of Maryland, Baltimore County, 1000 Hilltop Circle, Baltimore, MD 21250}
\altaffiltext{2}{NASA, Goddard Space Flight Center, 
Code 661, Greenbelt, MD 20771}

\begin{abstract}
The process responsible for the {\sl Chandra}-detected X-ray emission from  
the large scale jets of powerful quasars is a matter of ongoing debate. 
The two  main contenders are external Compton (EC) scattering off the cosmic 
microwave background (CMB)  photons (EC/CMB) and  synchrotron emission  
from a population of  electrons separate from those producing the radio - IR 
emission.
So far, no clear diagnostics have been presented to distinguish which of the 
two, if any, is the actual X-ray emission mechanism. Here we present  
such diagnostics based on a fundamental difference  between these two models: 
 the  production of synchrotron X-rays requires multi - TeV electrons,
while the EC/CMB model requires a cutoff in the electron energy distribution
(EED) below TeV energies.
This has significant implications for the $\gamma$-ray  emission predicted by 
these two  models, that can be tested through GeV and TeV observations
of the nearby  bright quasar 3C 273.  
We show how  existing
and future GeV and TeV observations can confirm
or refute one or both of the above hypotheses.

\end{abstract}


\section{Introduction\label{intro}}

The unexpected X-ray detection of the radio jet of quasar PKS 0637-752 by 
{\sl Chandra}
(Schwartz et al. 2000,  Chartas et al. 2000) ushered in  a new era in the 
study of extragalactic jets
(for a recent review see Harris \& Krawczynski 2006). 
The broadband spectral energy distribution (SED) of 
the jet knots exhibits a concave shape with
two spectral components, a low energy  synchrotron one peaking at 
$\nu\approx 10^{13}$  Hz  and cutting off at or below the optical regime, 
and a high energy one dominating the jet luminosity  and peaking 
beyond the {\sl Chandra} energy range.
Although some  X-ray synchrotron - self Compton (SSC) emission was anticipated,
the observed X-ray flux was found to be much higher than the SSC flux  
calculated  under equipartition conditions \citep{chartas00}. 
Tavecchio et al. (2000) and  Celotti, Ghisellini, \& Chiaberge (2001) 
proposed that the X-ray emission can be 
explained  as EC scattering off the CMB (EC/CMB), under the assumption that 
the  large scale jet is  relativistic with Lorentz factor $\Gamma\sim 10$ 
and points close to our line of sight, 
and   that its electron EED extends down 
to energies $\sim 10-100$ MeV,  significantly lower
than the $\sim 1-10$ GeV electron energies traced by  GHz synchrotron radio 
emission. Dozens  of other quasar jets, subsequently detected  by 
{\sl Chandra} (e.g. surveys by Sambruna et al. 2004, Marshall et al. 2005),  
share similar properties  and the EC/CMB interpretation of the X-ray extended
 jet  emission  became the interpretation of choice.

On  closer inspection, however, this interpretation
was shown to have  problems (see review by Harris \& Krawczynski 2006).
For example, because the cooling length of the low energy X-ray emitting 
electrons   is comparable to the jet length, the EC/CMB 
scenario cannot reproduce the kpc size X-ray knots and its low radiative 
efficiency leads to high, sometimes super - Eddington,  requirements on the 
jet kinetic power \citep{dermer04,uchiyama06}.
The problems encountered by the EC/CMB interpretation
lead many authors (e.g. Harris,  Mossman, \& Walker 2004, 
Kataoka \& Stawarz 2005, Hardcastle 2006, 
Jester et al. 2006, Uchiyama et al. 2006) 
to suggest that the X-rays are synchrotron emission from an additional EED
of multi-TeV electrons.
The synchrotron  interpretation is very well established in most low
 power  Fanaroff - Riley I  jets  \citep{fanaroff74}, where the apparently 
continuous SED  from radio to X-rays   strongly suggests a  single emission 
process (i.e. synchrotron) across this entire energy range 
(e.g. Hardcastle et al. 2002, Perlman \& Wilson 2005).
 
The two interpretations require electrons of
very different energies and, therefore, impose very different requirements
on  jet energetics and dynamics. For example, the high radiative efficiency
 of the synchrotron interpretation does not require high jet bulk lower 
Lorentz factors  (in agreement with 
arguments for $\Gamma \sim 2-3 $ from statistical radio  studies; 
Wardle \& Aaron 1997; Arshakian \& Longair 2004),
extreme lengths \citep{dermer04} or  super-Eddington jet power, 
as the EC/CMB does in several cases 
(Dermer \& Atoyan 2004; Jorstad \& Marscher 2004; Uchiyama et al. 2006).
Because these different requirements have a strong
impact on our understanding of quasar jets and of their effect on the 
host galaxy cluster, it is very important to devise ways to distinguish
between the two hypotheses.

Here we examine the GeV and TeV constraints imposed on these two 
different interpretations by existing and future GeV and TeV observations.
In particular, we show (\S \ref{3c273})  that for the nearby quasar 
3C 273, existing {\sl EGRET} and {\sl HESS} observations already
constrain the permitted parameter range of the two models, and future 
{\sl GLAST} and {\sl HESS} observations hold the potential of 
ruling one or, possibly,  both of them out.
We then (\S \ref{discussion}) discuss some additional points and caveats of 
this  work and draw our conclusions.

\section{What can we learn from GeV - TeV observations  of 3C 273? \label{3c273}}

3C 273 is the nearest ($z=0.158$) powerful quasar and, as such, 
it has been observed extensively. VLBI  observations reveal apparent 
superluminal  components in the pc scale jet with $u_{app}\approx 6-10$
 c (e.g. Unwin et al. 1985), constraining the pc-scale  jet orientation to 
$\theta <15^{\circ}$, and its Lorentz factor to $\Gamma>10 $. 
The actual angle of 3C 273's  jet to the line of sight cannot be very small, 
because, unlike  better aligned blazars, its
 optical spectrum exhibits a strong big blue bump 
(e.g. Courvoisier 1998, D'Elia, Padovani, \& Landt 2003).
 High dynamic range VLA  observations
 have only detected a one-sided large scale  jet,
with a jet to counter jet ratio  $R>10^4$ (e.g. 
Conway et al. 1993). 
  The ratio of the 
jet to counterjet flux is  
$R=(1+\beta\cos\theta)^{m+\alpha}/(1-\beta\cos\theta)^{m+\alpha}$,
where $\beta=u/c$ is the dimensionless jet speed, 
 $\alpha_r\approx 0.8$ is its radio spectral index, $m=2$  
for a  continuous jet, and $m=3$ for  discrete moving blobs
\citep{lind85}. 
We plot in figure \ref{beaming}  the constraints on the $\beta-\theta$ plane 
for the two cases. The  shaded area above each curve is excluded. Note
that this constraint does not depend on the nature of the jet X-ray emission.

Recent observations of 3C 273 with {\sl Chandra} and {\sl HST} \citep{jester06}
as well as {\sl Spitzer} \citep{uchiyama06}  show that the SED of each knot 
in the jet  is characterized by two  components, a low energy one having a 
cutoff 
above $\approx 5\times 10^{13}$ Hz, and a high energy one connecting the 
optical - UV and X-ray data.  For the purpose of this work we are interested 
in  the total jet SED, resulting by adding the fluxes of each knot, 
using the data from table 1 of Uchiyama et al. (2006).
We plot this in figure \ref{3c273sed} (diamonds),
fitted  with a two component (thick red and blue lines) analytical 
expression  following Uchiyama et al. (2006). 
Assuming an equipartition magnetic 
field  $B\delta\approx 10^{-4}$ G \citep{jester06}, 
where $\delta=1/\Gamma(1-\beta\cos \theta)$ is the  usual Doppler factor, 
implies that the  electrons producing the peak of the radio - IR component 
have energies  $\sim 0.2 $ TeV \citep{uchiyama06}, independent of the value of 
$\delta$  (Harris \& Krawczynski 2002; 
although see  Stawartz al. 2003 for a somewhat different result). If we assume
 that the optical to X-ray component is also synchrotron emission, 
the electrons of this
component  must reach energies at least up to $\sim 30 $ TeV, and the rising 
X-ray SED makes it plausible that the EED extends at least up to
 $\sim 100$ TeV.

Unavoidably, all these electrons will upscatter the CMB,
and the resulting EC/CMB  SED will be an exact copy of the synchrotron one,
 shifted  in frequency and luminosity. 
The frequency shift is:
\begin{equation}
{\nu_c \over \nu_s}={\nu_{CMB}\delta^2\gamma^2 \over e B \delta \gamma^2/(2\pi m_e c(1+z)) }=6.6 \times 10^8 \; \delta^2,
\end{equation}
where $\nu_c$ and $\nu_s$ are the observed EC and synchrotron frequencies 
emitted by  electrons of Lorentz factor  $\gamma$, $e$ and $m_e$ are the 
electron  charge and mass, $\nu_{CMB}=1.6 \times 10^{11}$ Hz is the CMB peak 
frequency at $z=0$, and   $B\delta= 10^{-4}$ G.
To derive the luminosity shift we note that the synchrotron observed $L_s$ and 
comoving $L_{s,com}$
luminosities are related through $L_s=L_{s,com} \delta^{3+m}$, while the EC
observed $L_c$ and 
comoving $L_{c,com}$
luminosities are related through $L_c=L_{c,com} \delta^{5+m}/\Gamma^2$, $m=0$ for a continuous flow and $m=1$ for discrete moving blobs.
  The  comoving luminosities are proportional
to the comoving energy densities, $U _{s,com}= B^2/(8\pi)$, 
$U_{c,com}= (4 /3)\Gamma^2 U_{CMB}(1+z)^4$ \citep{dermer94}, 
where $U_{CMB}=4.2 \times 10^{-13}$ erg cm$^{-3}$ 
is the CMB energy density at $z=0$. Using the above, we obtain
\begin{equation}
{L_c \over L_s }={32 \pi U_{CMB} (1+z)^4 \delta^4 \over 3 (B\delta)^2}=2.5 \times 10^{-3} \;\delta^4,\label{eq:L}
\end{equation}
where we again used  $B\delta= 10^{-4}$ G. 
The  two luminosities are  equal for $\delta=4.5$.

\subsection{GeV constraints\label{gev_const}}

\begin{figure}
\epsscale{0.8}
\plotone{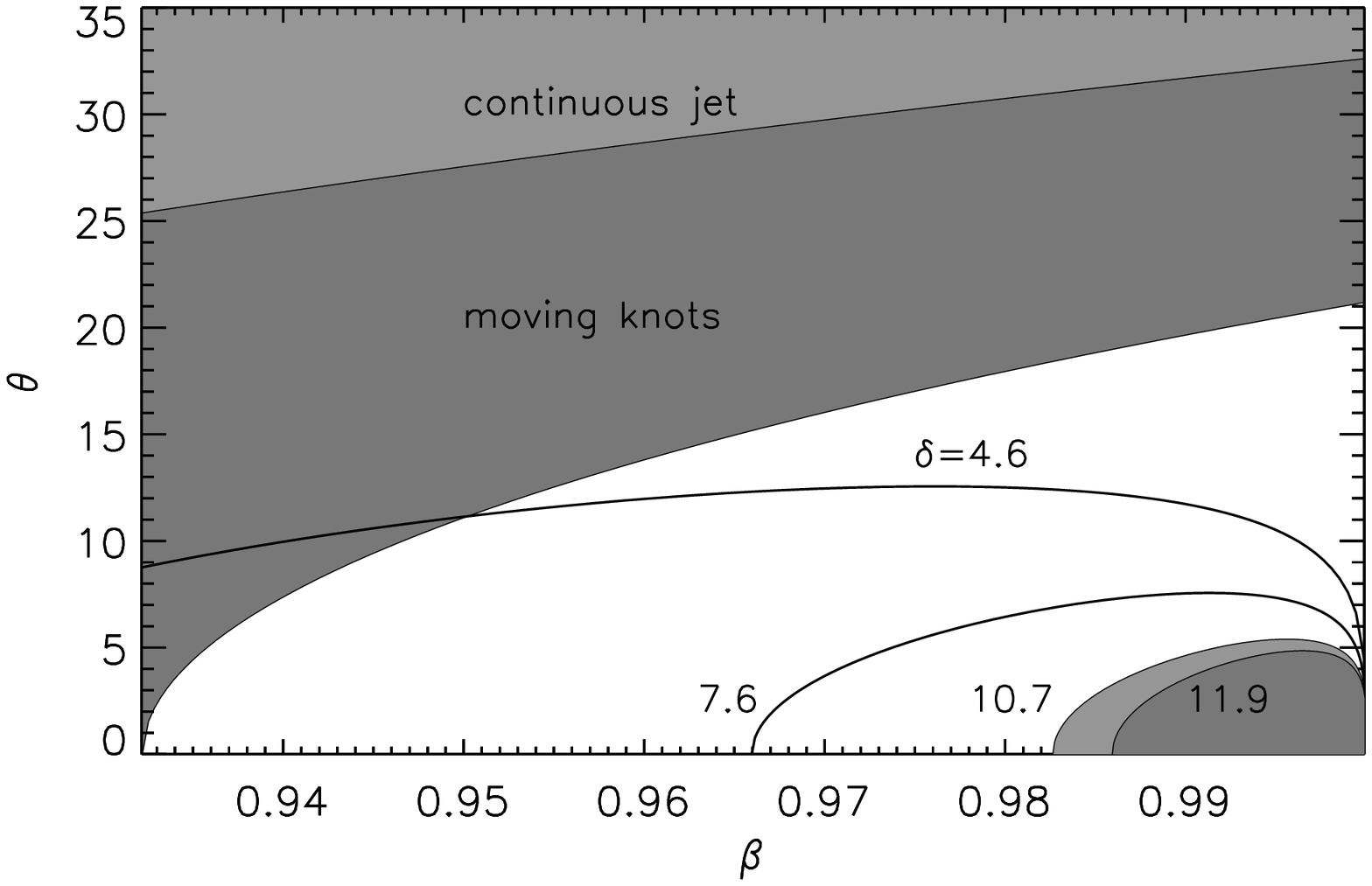}
\caption{The top shaded areas are the excluded areas  of the $\beta-\theta$ parameter space for the large jet of 3C 273, derived from constraints on the jet to counter jet ratio $R$. The four lines correspond to four different Doppler 
factors $\delta$. The bottom shaded areas, discussed in the text,
 represent  areas of the $\beta-\theta$ parameter space, excluded by existing
{\sl EGRET} and {\sl HESS} observations.}
\label{beaming}
\end{figure}

 The electrons producing the synchrotron IR peak  also produce a $\sim$ GeV 
peak from EC/CMB, the level of which scales with  $\delta^4$ 
(see eq. \ref{eq:L}). 
The existing {\sl EGRET} flux 
limit (see figure \ref{3c273sed}),  when the blazar is at its usual low state
\citep{vonmontigny97}, 
requires $\delta<11.9$, plotted as a 
dotted red line in figure \ref{3c273sed}. 
Note that this line, as well as the rest of the red lines of figure 
\ref{3c273sed}, does {\sl not} show the EC/CMB model a la Tavecchio et
al. (2000) and Celotti et al. (2001), which account for the X-rays
by postulating a low energy electron population, but rather  shows
the EC/CMB counterparts of the  {\sl observed} synchrotron emission 
for different Lorentz factors.
A line that corresponds to 
$\delta=11.9$ is also plotted in figure \ref{beaming} and the shaded 
region below  this line is excluded by the {\sl EGRET} limit.
{\sl Even this shallow limit on $\delta$, requires in the EC/CMB X-ray model
 either an $e-p$ super - Eddington jet  ( $L_{jet} \gtrsim 10^{49}$ 
erg s$^{-1}$)  or an $\sim$ Eddington power leptonic jet \citep{uchiyama06}.}

The fact  that 3C 273 was frequently below the {\sl EGRET} sensitivity limit,
makes it plausible that the  quiescent  GeV blazar emission
is significantly below  the {\sl EGRET} limit.
At the {\sl GLAST} sensitivity limit (see figure \ref{3c273sed}), 
a GeV non-detection during suitably  low blazar  states would set the  most 
 stringent limit of  $\delta< 4.7$, plotted as  a thin red solid line EC/CMB SED in figure \ref{3c273sed}. We also plot a $\delta=4.7$  line in figure \ref{beaming},
with the region below it corresponding to  $\delta> 4.7$.
The jet and blazar GeV emission of 3C 273 will be spatially unresolved by 
{\sl GLAST}.  To discriminate between the two, 
we note that the jet GeV emission would be   ({\sl i}) steady 
and  ({\sl ii}) much harder than the blazar one (photon index 1.8 as opposed 
to $\gtrsim 3.2$  for the  blazar at low states; von Montigny et al. 1997).
The jet emission, therefore,  should reveal itself as a steady hard plateau 
when the steep blazar  component drops below  the jet flux level.
As can  be seen in figure   \ref{beaming}, even for a {\sl GLAST} non - 
detection, there is a significant part of the $\beta-\theta$ space  
that is compatible with both the  $\delta< 4.7$  line and  $R$ constraints.
{\sl However, any limit on $\delta$ lower than 
the $\delta < 11.9$ limit of {\sl EGRET} further increases  the jet 
power requirements in the EC/CMB and  makes its application to 
3C 273 even more problematic.}

\subsection{TeV constraints\label{tev_const}}

\subsubsection{Constraints on the EC/CMB hypothesis}

 If the optical - X-ray emission is due to EC/CMB, then the highest energy 
electrons available are those emitting in the IR. As we discuss above,
these sub - TeV electrons will produce a GeV component, 
but are not energetic enough
to upscatter the CMB photons to TeV energies.
Therefore,  {\sl no TeV emission is expected in the EC/CMB model, and 
any hard, steady (i.e. of jet origin) TeV detection  from 3C 273, 
will rule out  the EC/CMB scenario}.    
Note that the potential contribution of the steep blazar spectrum 
({\sl EGRET} photon index 3.2 or steeper, 
when the blazar is  at its frequent low state; 
von Montigny et al. 1997) is expected to be significantly below the 50 hour
sensitivity limit of {\sl HESS}, with the possible exception of 
 high blazar  states when the {\sl EGRET} photon index 
hardens to 2.2 \citep{vonmontigny97}.

\subsubsection{Constraints on the synchrotron hypothesis}
What if the optical - X-ray emission is also synchrotron in nature?
In this case, the same  multi - TeV electrons that produce this emission will  
 also  produce a TeV  energy EC/CMB component at a level that
scales with $\delta^4$ (see eq. \ref{eq:L}). 
 Absorption 
by   the extragalactic background light (EBL) will reduce the TeV level and 
steepen its spectrum.
There is already a flux limit at 0.28 TeV  (plotted in figure \ref{3c273sed})
obtained  by a shallow  3.9 hours  
{\sl HESS}  observation  (Aharonian et al. 2005), assuming  an absorbed photon 
index of $3.0$.
To go from the emitted unabsorbed spectrum to the observed absorbed one, we  
adopt the baseline  analytical expression for the 0.2 - 2.0 TeV absorption 
optical depth of Stecker \& 
Scully (2006), scaled to agree at 0.28 TeV with  
the  EBL profile  (P0.45, as described by Aharonian et al. 2006)
estimated through {\sl HESS} observations  of  blazars 
H2356-309 and 1ES 101-232 at $z$ of 0.165 and 0.186 respectively.

We plot in  figure \ref{3c273sed} with short dashed lines
the SED for $\delta=10.7$. The red line is due to the electrons that 
produce the radio - IR emission, the blue line due to the electrons
producing the optical - X-ray emission in the synchrotron interpretation,
and the green line corresponds to the observed $0.2 - 2$  TeV flux, 
after the EBL absorption has been taken into account.      
As can be seen,  in the synchrotron X-ray scenario
 the  upper limit on $\delta$ has to be reduced to $\delta=10.7$
not to overproduce the existing  {\sl HESS} TeV upper limit. 
A $\delta=10.7$ line is also plotted in figure \ref{beaming}, with the area
shaded below it being excluded.

The {\sl HESS} flux limit that can  be achieved with longer exposure times 
can  been calculated from the 3.9 hour exposure limit,
assuming that the achievable flux limit scales with one over 
the square root of the exposure time. 
For a fixed synchrotron flux, and assuming equipartition, 
the EC/CMB flux $\propto \delta^{2+2\alpha_x}$, 
where  $\alpha_x$ is the X-ray spectral index.
The lowest $\delta$ jet that can be  detected for a given
exposure time, therefore, is
\begin{equation} 
\delta\propto t^{-1/(4+4\alpha_x)}.
\end{equation}
For a 50 hour {\sl HESS} observation,  $\delta=7.6$
(long dashed lines in figure \ref{3c273sed}, same interpretation as 
before). 
Therefore, {\sl for $7.6 <\delta< 10.7 $
{\sl HESS} will be able to detect the TeV emission predicted by the optical - 
X-ray synchrotron interpretation}.

\subsection{The combined constraints}

We present  now the combined constraints on the two hypotheses in terms
 of the  possible outcomes of {\sl GLAST} - {\sl HESS} observations: \\
($i$) We  detect neither  GeV  nor  TeV jet emission. This means that  
$\delta<4.7$. We can impose no constraints
 on the synchrotron scenario. However, {\sl the EC/CMB scenario 
is strongly disfavored} due to the extraordinary jet power requirements implied by the low value of $\delta$. \\
($ii$) We  detect the
jet GeV emission, at a low level that corresponds to  $4.7 < \delta <7.6$. 
We will not be able to detect the corresponding TeV 
emission, because it will be below the 50 hour {\sl HESS} sensitivity limit.
In this regime, again, we cannot constrain the synchrotron scenario, but 
{\sl the  low $\delta$ factors  still disfavor the EC/CMB scenario}. \\ 
($iii$) We  detect the
jet GeV emission, at a high level that corresponds to  $7.6 < \delta <10.7$.
Observationally, {\sl this is the most feasible case}, 
not only because the jet GeV flux 
is high,  but also because it is close to the {\sl EGRET} limit, and the 
conditions for detecting its hard steady emission during a low blazar 
state are the most favorable.
This is also the most interesting case, because TeV observations will 
confirm or reject the synchrotron model: The GeV flux  implies  a TeV 
flux in the synchrotron scenario, that is
 above the 50 hour {\sl HESS} sensitivity limit. 
{\sl If the anticipated TeV 
flux is observed, this will be a strong confirmation of the synchrotron model.
If not, then the synchrotron hypothesis for the X-ray emission is rejected.} \\

We close this section by discussing  the  lower limit $\gamma_{min}$ 
of the EED producing the radio-IR emission in the synchrotron X-ray hypothesis.
This is  constrained by the requirement that the low energy 
tail of the
EC/CMB GeV-peaking component does not produce any X-ray emission, or, 
more conservatively, does not dominate over the extended weak X-ray jet 
emission (the cross in figure \ref{3c273sed} represents the jet X-ray flux, 
excluding the bright knots A and B). 
This requires that 
$\gamma_{min}\gtrsim 1200 / \delta$ , and  becomes relevant  for 
$\delta \gtrsim 5 $: for $\gamma_{min}\lesssim 1200 $ and 
$ \delta \gtrsim 5  $ the X-ray emission produced through  EC/CMB 
 overproduces the, presumably synchrotron, extended X-ray emission.
The above constraint does not affect the GHz radio emission, which requires
 substantially  more 
energetic electrons ($\gamma\sim 2300 $).
A similar limit on $\gamma_{min}$  was found  by 
Jorstad \& Marscher (2004) for the quasar 0827+243.

\section{Conclusions and Discussion  \label{discussion}}

Although it is agreed that the radio-IR spectrum of powerful extended quasar 
jets is  synchrotron radiation, the  X-ray emission mechanism 
(synchrotron or EC/CMB)  
is a  matter of active debate. The synchrotron interpretation requires
a second population of electrons with energies at least up to $\sim 30-100$ 
TeV, while the   EC/CMB implies a cutoff of the EED at sub -  TeV energies.
In both cases, the EED responsible for the radio to IR synchrotron emission 
will produce a hard and steady GeV component. For the nearby 3C 273, 
the existing  {\sl EGRET} limits constrain the Doppler factor of the large 
scale jet to $\delta<11.9$ and {\sl GLAST} observations during the frequent, 
low level, steep  spectrum blazar states  can push this limit down to 
$\delta < 4.7$. 
{\sl Such low values for $\delta$ make application of the 
EC/CMB model problematic \citep{uchiyama06}. }
Because the Universe is practically transparent for GeV photons,
 {\sl {\sl GLAST} observations could  provide  upper limits 
on $\delta$ for sources 
at higher $z$.}

Recent {\sl HESS} detections of two blazars at  $z=0.165$ and $z=0.186$ 
 (Aharonian et al. 2006) suggest that the Universe is sufficiently 
transparent  to  allow the detection of  TeV emission from  
3C 273.  
Such emission is not expected in the EC/CMB model, because
the available electrons are not energetic 
enough to produce any TeV photons via CMB photon upscattering.
{\sl Detection, therefore, of any steady jet TeV emission will argue 
against  the EC/CMB model for the optical-X-ray emission of 3C 273. }
 The multi-TeV  electrons required from the synchrotron 
interpretation do produce such TeV emission at a level $\propto \delta^4$ 
(eq. \ref{eq:L}).
 Existing {\sl HESS} observations reject synchrotron models with 
$\delta>10.7$.  {\sl Coordinated {\sl GLAST} - TeV observations will be able 
to confirm or reject the synchrotron hypothesis  for $7.6 < \delta< 10.7$.}
 Note that, although a GeV detection close to the {\sl GLAST} sensitivity 
limit will only provide  limited spectral information,  
a detection  higher in flux by a factor $\sim 6-20$ at the important  
$7.6< \delta <10.7 $ regime, will provide sufficient 
counts to measure the GeV spectral slope and identify the source 
 of the GeV emission  (blazar or extended jet).

\begin{figure}
\epsscale{0.8}
\plotone{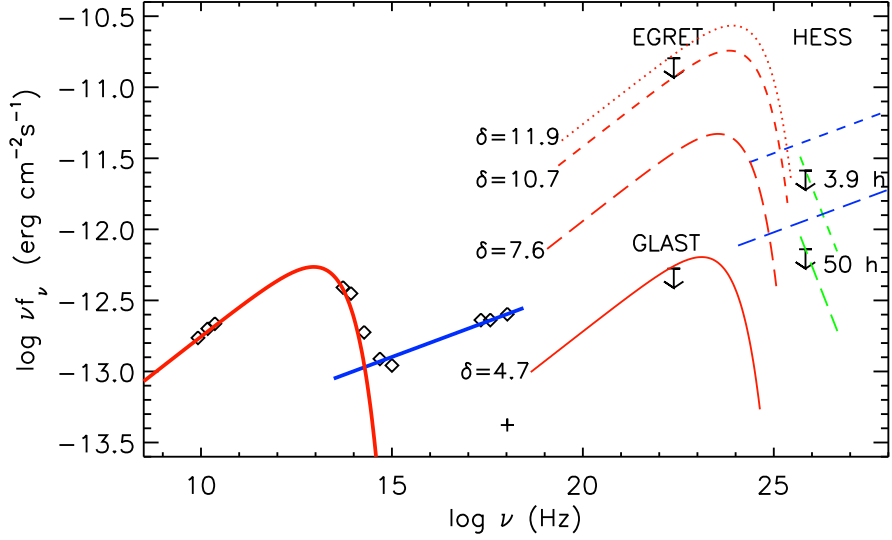}
\vspace{-40mm}
\caption{The radio to X-ray jet emission of 3C 273  (diamonds)   
fitted with a two component (thick red and blue lines) analytical 
expression  as in Uchiyama et al. (2006).  The cross represents the 
jet X-ray emission excluding the bright knots A and B.
The anticipated EC/CMB GeV emission due to the radio-IR emitting electrons 
 for $\delta=4.7$  (red line), 
 $\delta=7.6$  (long dashed red line),  
$\delta=10.7$  (short dashed red  line), and 
$\delta=11.9$  (dotted  red  line) are  also plotted, as well as the TeV 
emission, resulting in the synchrotron scenario from the optical - X-ray
emitting electrons for  $\delta=7.6$  (long dashed blue line for the source 
intrinsic emission, long dashed  green line for the EBL absorbed emission),  
and for $\delta=10.7$  (short  dashed blue line for the source 
intrinsic emission, short dashed  green line for the EBL absorbed emission).
The {\sl EGRET} upper limit and the {\sl GLAST}
sensitivity limits are shown, as well as a TeV flux upper limit obtained by shallow {\sl HESS} observations \citep{aharonian05}, together with the anticipated
50 hour {\sl HESS} sensitivity limit.}
\label{3c273sed}
\end{figure}

The actual TeV fluxes may be higher than our estimates.
In the synchrotron case, the  anticipated TeV flux  
increases if the  optical - X-ray emission is more beamed than the radio.
This could be the case if the X-rays come from a faster moving spine.
A thinner optical spine, compared to the radio jet width, has been observed 
in 3C 273 by Jester et al. (2005). Also, if the jet gradually decelerates
\citep{sambruna01,georganopoulos04}, it is possible that the X-ray emission 
that comes mostly from the first two knots A and B is more beamed than 
the radio-IR 
component that peaks further downstream. For a decelerating jet 
in the context of the EC/CMB model, the Doppler factor of the outer jet must be
sufficiently low, $\delta \lesssim 4 $, not to overproduce the extended 
 X-ray flux (cross in fig \ref{3c273sed}).

Our work is based on the assumption of energy equipartition between 
radiating electrons and magnetic field.
 As can be seen from eq. (\ref{eq:L}), if the actual conditions deviate
from equipartition with the  higher (lower) magnetic field, 
the anticipated GeV-TeV fluxes will have to be scaled 
down (up) by the square of the same factor and the observational limits on 
$\delta$ scaled up  (down) by the square  root of the same factor.
In the EC/CMB  interpretation the EED must 
extend down to $\gamma\sim 10-100$.
 Direct access  to this energy range requires low frequency, 
high resolution  radio telescopes such as the future Long Wave Array (LWA) 
\citep{harris06b}, that   will  be able to probe the radio emission produced
by the  low energy tail of the EED. 
We hope this work will motivate   TeV ({\sl HESS, VERITAS, MAGIC}) 
observations, which together with {\sl GLAST} GeV observations, 
hold the potential of identifying the X-ray energy emission process
of large scale quasar jets.

\acknowledgments
We thank the anonymous referee for suggestions that 
helped us improve this work.
M.G, E.S.P., and D.K. acknowledge support from NASA under  
LTSA grant NNG05-GD63DG and {\sl Chandra} theory grant TM6-7009A.

\clearpage

\clearpage

\end{document}